\documentclass[a4,twocolumn,prb,aps,showpacs,superscriptaddress,floatfix]{revtex4-1}
\usepackage[latin1]{inputenc}
\usepackage{bm}
\usepackage[usenames]{color}
\usepackage{multirow}
\usepackage{amssymb}
\usepackage{amsbsy}
\usepackage{amsmath}
\usepackage{stmaryrd}
\usepackage{graphicx}
\usepackage{epsfig}
\usepackage{placeins}
\usepackage{ulem}
\usepackage{hyperref}

\renewcommand{\vec}[1]{\bm{#1}}

\makeatletter

\begin{document}

\title{Exact asymptotic correlation functions of bilinear spin operators of the Heisenberg antiferromagnetic spin-$\frac{1}{2}$ chain}

\author {T. Vekua }
\affiliation{Institut f\"ur Theoretische Physik, Leibniz Universit\"at Hannover, 30167~Hannover, Germany}

\author{G. Sun}

\affiliation{Max-Planck-Institut f\"ur Physik komplexer Systeme, Dresden, Germany}

\begin{abstract}
Exact asymptotic expressions of the uniform parts of the two-point correlation functions of bilinear spin operators in the Heisenberg antiferromagnetic spin-$\frac{1}{2}$ chain are obtained. Apart from the algebraic decay, the logarithmic contribution is identified, and the numerical prefactor is determined. We also confirm numerically the multiplicative logarithmic correction of the staggered part of the bilinear spin operators  $\langle\langle  S^{a}_0S^{a}_{1}S^{b}_{r}S^{b}_{r+1}  \rangle\rangle=(-1)^rd/(r \ln^{\frac{3}{2}}r)  +(3\delta_{a,b}-1)  \ln^2r /(12 \pi^4 r^4)$, and estimate the numerical prefactor as $d\simeq 0.067$. The relevance of our results for ground state fidelity susceptibility at the Berezinskii-Kosterlitz-Thouless quantum phase transition points in one-dimensional systems is discussed at the end of our work.
\end{abstract}

\date{\today}
\pacs{64.70.Tg, 75.10.Jm, 75.10.Pq}


\maketitle

\section{ Introduction}
 The XXZ spin-$\frac{1}{2}$ chain is a paradigmatic one-dimensional quantum many-body system which can be studied using exact methods, while simultaneously describing the magnetic properties of real materials \cite{KM}. The Hamiltonian of the XXZ chain, written in terms of spin-$\frac{1}{2}$  matrices, reads,
\begin{equation} 
\label{ham-XXZ0}
\hat{H}_{XXZ}= J\sum_{r} \Big\{ S^{x}_{r}S^{x}_{r+1}
+S^{y}_{r}S^{y}_{r+1} +\lambda S^{z}_{r}S^{z}_{r+1}\Big\},
 \end{equation}
where $J$ is an exchange coupling, that will be assumed to be positive, and $\lambda$ is an anisotropy parameter. For $\lambda=1$, the XXZ chain reduces to the Heisenberg antiferromagnetic (AFM) chain.

Despite being exactly solvable, calculating correlation functions (objects that provide direct connection between theoretical calculations and experimental observations) from the microscopic XXZ model for $\lambda>-1$ is a formidable task, due to the complicated form of the wave functions \cite{Godin}. On the other hand, effective approaches, have allowed asymptotically exact calculation of the spin correlation functions in the gapless regime $-1<\lambda\le 1$ \cite{LZ,Affleck98,Lukyanov99,LT}.

Effective theory, describing the low-energy properties of the XXZ spin-1/2 chain for $-1<\lambda<1$, is given by the Gaussian model \cite{GNT},
\begin{equation} 
\label{ham-LL}
 H_G= \int  \mathcal{H}_G(r)\mathrm{d} r=\frac{v}{2}\int \mathrm{d}r \Big\{  (\partial_{r}\Phi(r))^{2}+\Pi^{2}(r)\Big\},
\end{equation}
where $\Phi$ is a real bosonic field with the compactification radius $R$, $\Phi=\Phi+2\pi R$, and $\Pi$ is its conjugate momentum, $[\Phi(r),\Pi(r')]=i\delta(r-r')$. Spin-wave velocity $v$ and $R$ are known analytically as functions of $\lambda$ from the exact solution of the model (\ref{ham-XXZ0}) \cite{JohnsonKrinskyMcCoy73}.

In this work, using an effective approach, we determine exact asymptotic expressions of the uniform parts of two-point correlation functions of bilinear spin operators in the Heisenberg AFM spin-$\frac{1}{2}$  chain. Apart from the algebraic decay, we identify the logarithmic contribution and determine the exact numerical prefactor. Our calculations are similar to the ones that were performed by Affleck for obtaining exact asymptotic correlation functions of single-spin operators in the Heisenberg antiferromagnetic chain \cite{Affleck98} by combining renormalization group (RG) improved perturbation theory with the exact asymptotic results of Lukyanov and Zamolodchikov \cite{LZ} conjectured for $-1<\lambda<1$. However, in the case of the correlation functions of the bilinear spins, there are various operators of effective field theory that contribute equally at the $SU(2)$ antiferromagnetic point, though the exact asymptotic expression is known for $-1<\lambda<1$ for the correlation function involving only one of them \cite{LT}. We have to use additional symmetry arguments to obtain exact asymptotic expression of the complete uniform parts of the bilinear spin correlation functions at $\lambda=1$.

We also identify numerically the multiplicative logarithmic contribution of the staggered (leading) parts of the bilinear spin correlation functions, consistent with analytical prediction \cite{Affleck+89}, and estimate the numerical prefactor. 

\section{Single-spin correlation functions} 
Asymptotic expressions of the single-spin correlation functions of the XXZ spin-$\frac{1}{2}$  chain, $ G^a_r=\langle S^a_{0}S^a_r\rangle$, where $a=x,y,z$ and no summation with repeated indices is implied in this work, are known exactly in the gapless phase, including the numerical prefactors \cite{LZ,Lukyanov99,LT}
\begin{eqnarray}
\label{ampl}
\! G^x_r=\!\frac{(-1)^r\!A_0^x}{r^{\eta}}-\frac{A_1^x}{r^{\eta+1/\eta}},\, G^z_r=\!\frac{(-1)^r\!A_1^z}{r^{1/\eta}}-\frac{1}{4\eta \pi^2 r^2},
\end{eqnarray} 
where $\eta=1-({\arccos{\lambda}})/{\pi}=2\pi R^2$ and $0 \le \eta \le 1$ for $-1\le \lambda \le 1$.

 These amplitudes appearing in Eqs. (\ref{ampl}) have been checked numerically \cite{HF98}. Amplitudes $A_0^x$ and $A_1^z$ diverge in the isotropic AFM limit, $\lambda \to 1$, since in this limit mapping of the spin-$\frac{1}{2}$  chain to the Gaussian model becomes singular due to the marginally irrelevant (cosine) term with the scaling dimension $2/\eta$ occurring in the low-energy effective theory from the ``spin umklapp'' processes \cite{Haldane,BE,dN,Affleck85}. 

Effective theory description of the XXZ spin-$\frac{1}{2}$ chain, for $\lambda\to 1$, necessarily contains terms beyond the Gaussian model, 
\begin{equation}
\label{effectiveH}
H_{eff}=\int \mathrm{d}r\mathcal{H}_{eff}=\int \mathrm{d}r[\mathcal{H}_G(r)+ \mathcal{H'}_1(r)+\mathcal{H'}_2(r)],
\end{equation}
 where $  \mathcal{H'}_1=\frac{\pi g^0_{||}}{\sqrt{3}}  [  (\partial_r \Phi)^2-\Pi^2 ] $ and $\mathcal{H'}_2=\frac{\pi g^0_{\bot}}{\sqrt{3}} \cos{\sqrt{8\pi }\Phi}$. 
The running coupling constants ${\vec g}=(g_{||},g_{\bot})$, with the bare values ${\vec g^0}= (g^0_{||}, g^0_{\bot})$, are governed by the Kosterlitz-Thouless \cite{Kosterlitz74} RG equations,
\begin{eqnarray}
\label{KTRG}
\beta_{||}=\dot  g_{||}=-4\pi g^2_{\bot}/\sqrt{3},\,\, \beta_{\bot}=\dot  g_{\bot}=-4\pi g_{\bot}g_{||}/\sqrt{3},
\end{eqnarray}
where the dot indicates a derivative with respect to the RG scale $l=\ln{\tilde r}$ and $\pi \tilde r^{-1}$ is the running ultraviolet cutoff. At the $SU(2)$ AFM point $g_{||}= g_{\bot}=g$, and the exact expression of amplitudes of the asymptotic correlation functions of spin operators were derived \cite{Affleck98,Lukyanov98} by combining the expressions of $A_0^x$ \cite{LZ} and $A_1^z$ \cite{Lukyanov99} for $\lambda\to 1_-$ with RG improved perturbation theory \cite{Cardy86,Affleck+89},
\begin{equation}
\label{su2}
\langle S^{a}_{0}S^{b}_r \rangle=(-1)^r\delta^{ab}\frac{\sqrt{\ln r}}{(2\pi)^{3/2}r}-\frac{\delta^{ab}}{4\pi^2 r^2},
\end{equation}
where $a,b=x,y,z$.

Prior to the analytical works \cite{Affleck98,Lukyanov98}, the numerical prefactor of the staggered term in Eq. (\ref{su2}) has been estimated by numerical simulations \cite{KomaMizukoshi} as $0.065$, which is close to the exact value $1/(2\pi)^{3/2}\simeq 0.0635$. 

\section{Bilinear spin correlation functions} 
We will generalize the approach leading to the exact assymptotic expressions of single-spin correlation functions at the Heisenberg AFM point \cite{Affleck98,Lukyanov98}, Eq. (\ref{su2}), for the calculation of the uniform part of the correlation function of bilinear spin operators, $S_{r}^aS_{r+1}^a$.

Let us first address the correlation function of the bilinear in $S^x$ operator in the gapless region in the vicinity of (but not directly at) the $SU(2)$ AFM point. Up to the subleading corrections we have
\begin{eqnarray}
\label{ccor}
\!\!\!\! \langle  S_{0}^xS_{1}^x S_{r}^xS_{r+1}^x \rangle\!=\!B_0+ \frac{(-1)^r\! B_1}{r^{1/\eta}}  +  \frac{B_2}{r^{4\eta}}+   \frac{B_3}{r^{4/\eta}}+\frac{B_4}{r^4}.
\end{eqnarray}
The $B_l$'s, for $l\ge 2$, are amplitudes of the correlation functions of the following $\hat O^{x}_l$ operators, 
\begin{eqnarray}
\label{prop}
 \hat O^{x}_2 &\sim& \cos \sqrt{8 \pi \eta} \Theta, \quad \hat O^{x}_3 \sim \cos \sqrt{8 \pi/ \eta} \Phi \nonumber \\
\hat O^{x}_4 &\sim & (\partial_r \Phi)^2 +  \beta^{x}_{\eta} (\partial_r \Theta)^2, \quad \partial_r\Theta= \Pi
\end{eqnarray}
and $\beta^{x}_{\eta}$ is, similarly to other proportionality coefficients in (\ref{prop}), an $\eta$-dependent factor such that 
\begin{equation}
\label{correctidenx}
\sum_r S^{x}_{r}S^{x}_{r+1} \to  \int \mathrm{d} r  \{ \sqrt{B_0}+  \hat O^{x}_2  +   \hat O^{x}_3+ \hat O^{x}_4 \}.
\end{equation}
The scaling dimension of $ \hat O^{x}_2$ is $2\eta$,  while those of $\hat O^{x}_3$ and $\hat O^{x}_4$ are $2/\eta$ and $2$, respectively. In the limit of the $SU(2)$ AFM point $\eta\to 1$ and all of them become marginal.

The constant term in Eq. (\ref{ccor}) can be easily fixed due to the translational symmetry, $ B_0=\frac{1}{4}(e_0-\lambda \partial e_0/\partial \lambda)^2$, where $e_0$ is the ground state energy density known exactly (together with its dependence on $\lambda$) from the Bethe ansatz. However we will be interested in the following with 
the reduced correlation function,  
\begin{equation}
\label{totalbl}
\langle  \langle S_{0}^xS_{1}^x S_{r}^xS_{r+1}^x   \rangle \rangle\! = \! G^{x,x}_r  \!= (-1)^rG^{x,x}_{s}(r)+G^{x,x}_{u}(r).
\end{equation}
Namely, the uniform part of the above reduced correlation function is the main quantity of our interest, 
\begin{eqnarray}
\label{bl}
G^{x,x}_{u}(r)&=& G^{x,x}_{B_2}(r)+G^{x,x}_{B_3}(r)+ G^{x,x}_{B_4}(r)\nonumber\\
 &=& \frac{B_2}{r^{4\eta}}+ \frac{B_3}{r^{4/\eta}}+  \frac{B_4}{r^{4}}.
\end{eqnarray}
 For the XXZ chain, for $-1<\lambda<1$, the exact expression of $B_2$ amplitude has been obtained \cite{LT}, 
\begin{equation}
\label{LT}
B_2= \frac{[\Gamma(\eta)]^4}{2^{3+4\eta}\pi^{2+2\eta} (1-\eta)^2}\left[   \frac{ \Gamma(\frac{1}{2-2\eta})}{\Gamma(\frac{\eta}{2-2\eta})} \right]^{4-4\eta}
\end{equation}
 and confirmed numerically away from the $SU(2)$ points \cite{HF04}. In Appendix A we provide details of calculating $B_2$, confirming expression (\ref{LT}). However, when $\eta\to 1$, the expression for $B_2$ is only valid for evaluating correlations Eq.(\ref{bl}) at exponentially large distances, $r\gg e^{1/(1-\eta)}$. In the limit of the $SU(2)$ AFM point we apply RG improved perturbation theory \cite{Affleck98}. We note that $G^{x,x}_{B_2}$ obeys the following RG equation,
\begin{equation}
\label{RGEQ}
\Big(\frac{\partial}  {\partial \ln r} +\sum_{j=\{||,\bot \}} \beta_j\frac{ \partial}{ \partial g_j}+2\gamma_{B_2}(\vec g)\Big) G^{x,x}_{B_2}(r,\vec g) =0,
\end{equation}
where $\beta_j$ are beta functions presented in Eq. (\ref{KTRG}) and $\gamma_{B_2}(\vec g)=2-4\pi g_{||}/\sqrt{3}$ is the anomalous dimension of the $\hat O^x_2$ operator, calculated in Appendix B. This allows us to follow the approach \cite{Affleck98,Lukyanov98} that led to the exact expression of the single-spin correlation function Eq. (\ref{su2}). Solving the RG Eq. (\ref{RGEQ}) and integrating over $\gamma_{B_2}(\ln{r})$ in the solution, as shown explicitly in Appendix C, gives for $1-\eta \ll 1$ the following behavior over an intermediate range $1\ll \ln{r}\ll1/(1-\eta)$: $G^{x,x}_{B_2}(r)\simeq {B_2 (4(1-\eta)\ln{r})^2}/{r^4}$. Then taking the limit $\eta\to 1$ and using the limiting expression of the amplitude in Eq. (\ref{LT}), $B_2\to {1}/({2^7\pi^4(1-\eta)^2})$, we obtain the following exact asymptotic expression for $\ln r \gg 1$,
\begin{equation}
\label{ee}
G^{x,x}_{B_2}(r)= \frac{1}{8\pi^4}  \frac{\ln^2r}{ r^4}.
\end{equation}

Let us consider now the mixed correlation function of bilinear spin operators at the Heisenberg AFM point, 
\begin{equation}
\langle  \langle S^{a}_0S^{a}_{1}S^{b}_{r}S^{b}_{r+1} \rangle \rangle=G^{a,b}_r=(-1)^r G^{a,b}_{s}(r)  +G^{a,b}_{u}(r),
\end{equation}
for $a \neq b$. Using bosonization \cite{GNT} one can show that the leading staggered part of the mixed correlation function $G^{a,b}_{s}(r)$ behaves identically to $G^{a,a}_{s}(r)$ \cite{comment0}. To study the long-distance asymptotics of the uniform part of the mixed bilinear correlation function, $G^{a,b}_{u}(r)$ for $\ln{r}\gg 1$, it is useful to look at the correlation function of the Hamiltonian density,
\begin{equation}
\langle \langle   ( {\bf S}_0{\bf S}_1 )( {\bf S}_r{\bf S}_{r+1})  \rangle \rangle=(-1)^rG^E_s(r)+G^E_u(r),
\end{equation}
 where $ G^E_s(r)=9 G^{a,a}_{s}(r)+\cdots$, and we will use the important property that the uniform part $G^E_u(r)$ can not contain multiplicative logarithmic corrections due to energy conservation \cite{comment}. Since energy density does not pick up anomalous dimension due to marginally irrelevant perturbations, the correlation function of energy density behaves similarly to the correlation function of the energy-momentum tensor of the unperturbed conformally invariant Gaussian or Wess-Zumino model \cite{comment01}, thus $G^E_u(r)\sim 1/r^4$. This means that logarithmic contributions, such as in Eq. (\ref{ee}), all must be canceled by the mixed terms. Hence, at the $SU(2)$ AFM point, for the leading behavior of the uniform parts of the bilinear spin correlation functions, we obtain,
\begin{equation}
\label{ee1}
2G^{a,b\neq a}_{u}(r)= - G^{a, a}_{u}(r).
\end{equation}
Let us rewrite the relation (\ref{correctidenx}) in the following way: $\sum_r S^{x}_{r}S^{x}_{r+1} \to  \int \mathrm{d} r  \{ \sqrt{B_0}+ \hat O^{x}_2+ \hat W \}$, by grouping two operators into one $\hat W=\hat O^{x}_3+\hat O^{x}_4$. Then, from bosonization, it follows that
$\sum_r S^{y}_{r}S^{y}_{r+1} \to  \int \mathrm{d} r  \{ \sqrt{B_0} - \hat O^{x}_2+ \hat W \}$. For $\lambda \to 1$, using Eq. (\ref{ee1}) for the case of $a=x$ and $b=y$, we obtain $3\langle\langle \hat W(0) \hat W(r)\rangle \rangle =\langle\langle \hat O^{x}_2(0) \hat O^{x}_2(r)\rangle \rangle$
and thus,
\begin{equation}
\label{mainresult}
G^{a,a}_u(r)=\frac{4}{3}G^{x,x}_{B_2}(r)=\frac{1}{6\pi^4}  \frac{\ln^2r}{ r^4}.
\end{equation}
This exact asymptotic expression is our main result.

\begin{figure}
\includegraphics[width=7.5cm]{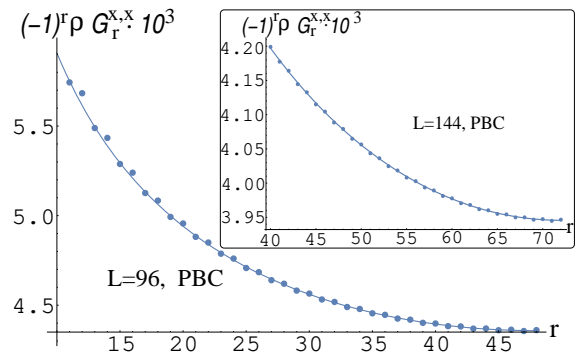}
\caption{Bilinear spin correlation function of the Heisenberg spin-$\frac{1}{2}$  AFM chain $(-1)^r\rho G^{a,a}_r\cdot 10^3$, where we have introduced the cord distance on the circle $\rho= L/\pi\sin(r\pi/L)$. Data shown are for $L=96$ spins with periodic boundary conditions (PBC) and are obtained by keeping typically $m=1500$ states in DMRG simulations. Bullets indicate numerical data \cite{comment1} and continuous line is the curve, $10^3d/\ln^{\frac{3}{2}}(c\rho)$, with $d\simeq 0.067$ and $c\simeq 16$ coefficients obtained by fits of analytical curve to numerical data for $r>10$. Inset shows $(-1)^r\rho G^{a,a}_r\cdot 10^3$, for $L=144$ spins and for distances, $40 \le r\le 72$. Fitting to the data for $L=144$ sites gives similar estimates for $d$ and $c$.}
\label{fig:pd}
\end{figure}

\section{Numerical results}
In the remaining part we will present a numerical check of Eq. (\ref{mainresult}) based on our results obtained from the density matrix renormalization group
(DMRG) method \cite{White,Uli} implemented for systems with periodic boundary conditions. 

Directly from the computation of the reduced bilinear spin correlation function Eq. (\ref{totalbl}) it is hardly possible to analyze the space dependence of its uniform part. The reason is that the reduced correlation function $G^{a,a}_{r}$, is strongly dominated by the leading term, its staggered part, which is expected to behave as
$G^{a,a}_{s}(r)\sim  1/(r \ln^{\frac{3}{2}}r)$ \cite{Affleck+89}, and such decay is much slower than that of the uniform part ($\sim \ln^2r/r^4$). We have performed numerical simulations of $G^{a,a}_{r}$ for different system sizes, ranging from $L=24$ (Lanczos) to $L=48, 96$ and $L=144$ sites (DMRG), assuming periodic boundary conditions. In Fig. 1 we present the behavior of the reduced correlation function of the bilinear spin operators for the Heisenberg spin-$\frac{1}{2}$ AFM chain with $L=96$ and $L=144$ sites. We use conformal mapping of an infinite 2-dimensional plane on a cylinder \cite{comment01} with finite circumference in the spatial direction to compare the analytic results for the thermodynamic limit with finite-size calculations for the systems with periodic boundary conditions. This implies that the distances are replaced by the chord distances on the circle, $\rho= L/\pi\sin(r\pi/L)$.

On the other hand, in the difference $G^{a,a}_r- G^{a,b}_r$, $b\neq a$, the leading oscillatory terms cancel \cite{comment0} and from this quantity and Eq. (\ref{ee1}) we can obtain the desired uniform part of the correlation function $G^{a,a}_{u}(r)$. In Fig. 2 we plot numerical data for $r^4(G^{a,a}_r-G^{a,b}_r)$, which includes both uniform and staggered components. For the uniform component our analytical result is $\ln^2r/(4\pi^4)$, following from Eq. (\ref{mainresult}) and the relation $G^{a,s}_r-G^{a,b}_r=\frac{3}{2}G^{a,a}_{u}(r)+\cdots$, where dots indicate sub-leading contribution. We will calculate the leading oscillatory contribution in
\begin{equation*}
\!\!2(G^{x,x}_r\!\!-\!G^{x,y}_r)=\! \langle \! { (S^x_0S^x_1-\!S^y_0S^y_1 )(S^x_rS^x_{r+1}-\!S^y_rS^y_{r+1} )} \! \rangle.
\end{equation*}

\begin{figure}
\includegraphics[width=8.5cm]{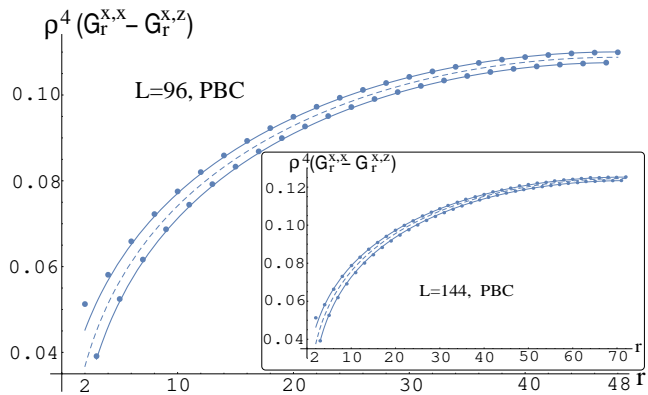}
\caption{Difference of bilinear spin correlation functions of the Heisenberg spin-$\frac{1}{2}$ AFM chain $\rho^4(G^{a,a}_r-G^{a,b}_r)$ for $a\neq  b$ and for $L=96$ sites chain with periodic boundary conditions (PBC). Bullets indicate numerical data and continuous lines are analytical curves, $\ln^2(c_0\rho)/(4\pi^4)\pm \tilde c{\ln^{\frac{3}{2}}(c_0 \rho)}/\rho$, where $+$ sign corresponds to even $r$ data and $-$ sign to odd $r$ data. Constant $c_0\simeq 22$ and $\tilde c$ is fitted to $\tilde c\simeq 0.0023$. Dashed line is analytical result without taking into account leading oscillatory contribution. Inset shows $L=144$ sites case with $c_0\simeq 23$ and the same value of $\tilde c\simeq 0.0023$.
}
\label{fig:pd}
\end{figure}

In bosonization $ (-1)^r(S^x_rS^x_{r+1}-S^y_rS^y_{r+1})\to \hat D(r)$, where $\hat D(r) \sim e^{i{\sqrt{8\pi}\Theta(r) }}\cos{\sqrt{2 \pi} \Phi(r)}+H.c.$, up to sub-leading contributions.
The anomalous dimension of $\hat D$ is $\gamma_D=5/2-\sqrt{3}\pi g/2$, giving 
$
\langle \hat D(0)\hat D(r)  \rangle \sim   {{\ln^{\frac{3}{2}} r}}/{r^5}.
$

Hence, including the leading oscillatory contribution in $G^{a,a}_r-G^{a,b}_r$, we obtain
\begin{equation}
\label{FE}
G^{a,a}_r\!-G^{a,b}_r=(1-\delta_{a,b}) \left[ \frac{ \ln^2{r}}{4\pi^4 r^4}+\tilde c (-1)^r  \frac{{\ln^{\frac{3}{2}} r}}{r^5} \right],
\end{equation}
where $\tilde c$ is a numerical constant estimated from fitting to DMRG data. We present in Fig. 2 comparison of our analytical curves, obtained separately for even and odd $r$ from Eq. (\ref{FE}), with our numerical data.

\section{Relevance for fidelity susceptibility} 
The asymptotically exact expression of the uniform part of the correlation function of the bilinear spin operators at $\lambda=1$, Eq. (\ref{mainresult}), confirms our previous work \cite{Vekua1}, showing that computing the ground state fidelity susceptibility of the XXZ spin-$\frac{1}{2}$ chain by the effective Gaussian model gives a qualitatively wrong result at the Berezinskii-Kosterlitz-Thouless quantum phase transition point. Moreover, our approach allows us to explicitly follow the steps on how the divergence in fidelity susceptibility at the Berezinskii-Kosterlitz-Thouless transition point arises in the thermodynamic limit due to the singular nature of the mapping of the Heisenberg spin-$\frac{1}{2}$ AFM chain on the Gaussian model and is not a property of either the microscopic or effective models. When taking into account marginally irrelevant corrections to the effective Gaussian model and resumming perturbation series with the help of the RG, the spurious divergence of fidelity susceptibility disappears, as explained in Appendix D.

\section{Summary} Using an effective field-theory approach, exact asymptotic expressions of the uniform parts of the biliniar spin correlation functions of the Heisenberg antiferromagnetic spin-$\frac{1}{2}$  chain, $G^{a,a}_u(r)$ and $G^{a,b}_u(r)$,  have been computed. We have checked numerically analytical results and also estimated the numerical prefactor in front of the staggered part of the bilinear spin correlation function and identified the logarithmic contribution in accordance with the previous analytical investigations \cite{Affleck+89}. 

As a by-product, our studies confirm the finiteness of the ground state fidelity susceptibility at the Berezinskii-Kosterlitz-Thouless quantum phase transition points in one-dimensional systems.

\section{Acknowledgments} 
This work has been supported by DFG Research Training Group (Graduiertenkolleg) 1729 and Center for Quantum Engeneering and Space-Time research (QUEST). We thank A. Furusaki, T. Hikihara, A. K. Kolezhuk, S. Lukyanov, and M. Oshikawa for helpful discussions.

\section{Appendix}

\subsection{Calculating constant $B_2$ for $-1<\lambda<1$}
We provide details of calculating the exact expression of constant $B_2$ appearing in Eq. (12) in the main text.
For this we introduce the Hamiltonian of the fully anisotropic XYZ spin-$\frac{1}{2}$ chain, 
\begin{equation} 
\label{ham-XXZ}
\hat{H}_{XYZ}= \sum_{r} \Big\{J_x S^{x}_{r}S^{x}_{r+1}
+J_yS^{y}_{r}S^{y}_{r+1} + J_z  S^{z}_{r}S^{z}_{r+1}\Big\}.
 \end{equation}
We will assume $|J_z|<J_y\le J_x$. Denoting  $\Delta=J_z/J$, $J=(J_x+J_y)/2$, and $\gamma= (J_x-J_y)/(2J)$, we rewrite the Hamiltonian of the XYZ chain as follows,
\begin{equation*} 
\hat{H}_{XYZ}= J \sum_{r} \Big\{ (1+\gamma)S^{x}_{r}S^{x}_{r+1}
+ (1-\gamma)S^{y}_{r}S^{y}_{r+1} + \Delta  S^{z}_{r}S^{z}_{r+1}\Big\}.
 \end{equation*}

We will put $\hbar=1$ and measure energy in units of $J$.
Spin-wave velocity for the gapless, $\gamma=0$, case is
\begin{equation}
\label{velo}
v=\frac{Jr_0 \sin{(\pi \eta)}}{2(1-\eta)},
\end{equation}
where $r_0$ is the lattice constant and $\eta=1-({\arccos{\Delta}})/{\pi}$.
It is convenient to fix the spin-wave velocity equal to unity (hence also make dimensionless) for $\gamma=0$, independently of $\Delta$. For this, we will fix
\begin{equation}
\label{fixation}
Jr_0=2\frac{1-\eta}{\sin{(\pi\eta)}}.
\end{equation}

In the following we will use the exact solution of the XYZ chain \cite{Baxter82,JohnsonKrinskyMcCoy73}. In particular we will be interested in the limit $\gamma\to 0$, and take the so called scaling limit of the XYZ chain, where the spin gap behaves as \cite{Luther76}
\begin{eqnarray}
\label{Gap}
M_{XYZ}&=&2 J_x \frac{\sin({\pi \eta})}{ 1-\eta} \left( \frac{\sqrt{J_x^2-J_y^2}  }{4\sqrt{J_x^2-J^2_z}} \right)^{\frac{1}{1-\eta}} \nonumber\\
&\simeq& 2 J \frac{\sin({\pi \eta})}{ 1-\eta} \left( \frac{\gamma}{4(1-\Delta^2)} \right)^{\frac{1}{2(1-\eta)}} \nonumber\\
&=&\frac{4}{r_0} \left( \frac{\gamma}{4\sin^2(\pi \eta)} \right)^{\frac{1}{2(1-\eta)}}\nonumber\\
&=&4\left( \frac{\gamma_r}{4\sin^2(\pi \eta)} \right)^{\frac{1}{2(1-\eta)}},
\end{eqnarray} 
where $\gamma_r= \gamma r_0^{-2(1-\eta)}$ and to arrive from the second to the third line we used Eq. (\ref{fixation}). The scaling limit is a continuous limit of the lattice model, $r_0\to 0$, with additional requirements: the velocity Eq. (\ref{velo}) stays equal to unity and the gap Eq. (\ref{Gap}) stays constant; hence $\gamma \to 0$ so that $\gamma_r=const$. In this limit, the effective theory describing the XYZ chain is a massive relativistic sine-Gordon model \cite{Luther76},
\begin{equation} 
\label{ham-sG}
 A_{sG}=\frac{1}{2} \int \mathrm{d^2} {\vec r}  ( \partial_{\mu} \Theta)^2   -2\mu \int \mathrm{d^2} {\vec r} \cos{ \sqrt{8 \pi \eta } \Theta}  .
\end{equation}
To give explicit meaning to $\mu$ one has to specify normalization of fields. We will follow the approach developed by Zamolodchikov \cite{Zamolodchikov}, where the dimension of field $[\cos{ \sqrt{8 \pi \eta } \Theta}]=r_0^{-2\eta}$ and the fields are normalized as follows at short distances, where perturbation is irrelevant \cite{comment}:
\begin{equation}
\label{SDN}
\lim_{r\to 0}\langle  \cos{ \sqrt{8 \pi \eta } \Theta(0)}  \cos{ \sqrt{8 \pi \eta } \Theta(r)}  \rangle= \frac{1}{2}\frac{1}{r^{4\eta}}.
\end{equation}

Explicit connection between the coupling constant $\mu$ and the soliton mass of the sine-Gordon model was obtained \cite{Zamolodchikov} by using the Bethe ansatz integrability of the sine-Gordon model in external uniform gauge field, with amplitude $A$, coupled to the conserved current \cite{JNW} and viewing the same model as a conformal field theory (Gaussian model for $\mu=0$) perturbed by a cosine term. In the Bethe ansatz approach, the ground-state energy of the quantum sine-Gordon model in strong external field, $[E_0(A,M_{s-G})-E_0(0,M_{s-G})]/A^2$, can be expanded in a dimensionless parameter, the ratio of the soliton mass to the field amplitude $M_{s-G}/A$. On the other hand, when viewing the sine-Gordon model in strong external field as a perturbation of conformal field theory with cosine field, the ground state energy, $[E_0(A,\mu)]-E_0(A,0)]/A^2$, can be expanded in the powers of dimensionless parameter $\mu/A^{2/(p+1)}$. Note that without strong external field, ground-state energy can not be perturbatively expanded in $\mu$, due to the infrared divergent integrals characteristic of the relevant cosine term. Matching the two ground-state energies in the first nontrivial power of $A$ gives
\begin{equation}
\label{MsG}
M_{sG}=\left( \frac{ \mu}{ \kappa(\eta)}  \right)^{\frac{1}{2(1-\eta)}},
\end{equation}
where the dimensionless parameter $\kappa$ is called the $\lambda -M$ ratio (proportionality constant between ultraviolet and asymptotic scales) and is given by \cite{Zamolodchikov} 
\begin{equation}
\kappa(\eta)=\frac{1}{\pi}\frac{\Gamma(\eta)}{\Gamma(1-\eta)}\left(  \frac{\sqrt{\pi}  \Gamma(\frac{1}{2(1-\eta)}) }{2\Gamma(\frac{\eta}{2(1-\eta)}) }  \right)^{2(1-\eta)}.
\end{equation}

We wish to determine a proportionality constant $\alpha$,
\begin{equation}
\label{mG}
2\mu=- \alpha \gamma_r J r_0
\end{equation}
in order to obtain a precise value of the constant in the operator identification of bilinear spin operators in the scaling limit,
\begin{eqnarray}
&&J \gamma \sum_r(S^{x}_{r}S^{x}_{r+1}-S^{y}_{r}S^{y}_{r+1} )= \frac{J \gamma}{r_0} \sum_r (S^{x}_{r}S^{x}_{r+1}-S^{y}_{r}S^{y}_{r+1} )r_0\nonumber\\
&&=\frac{J \gamma}{r_0} \frac{r_0^{2\eta}}{a^{2\eta}} \sum_r(S^{x}_{r}S^{x}_{r+1}-S^{y}_{r}S^{y}_{r+1} )r_0\nonumber\\
&&=\alpha {J \gamma r_0^{2\eta-1}  }   \int \mathrm{d} r  \cos{ \sqrt{8 \pi \eta } \Theta} \nonumber\\
&&=\alpha {Jr_0 \gamma_r  }   \int \mathrm{d} r  \cos{ \sqrt{8 \pi \eta } \Theta}=-2\mu  \int \mathrm{d} r  \cos{ \sqrt{8 \pi \eta } \Theta},
\end{eqnarray}
where $  \sum_r(S^{x}_{r}S^{x}_{r+1}-S^{y}_{r}S^{y}_{r+1} )r_0/r_0^{2\eta}\to \alpha \int \mathrm{d}r\cos{ \sqrt{8 \pi \eta } \Theta}$.

With the help of Eqs. (\ref{MsG}) and (\ref{mG}) we express the sine-Gordon mass as
\begin{equation}
M_{sG}= \left( \frac{ -\alpha \gamma_r Jr_0}{ 2\kappa(\eta)}  \right)^{\frac{1}{2(1-\eta)}}.
\end{equation}

Equating $M_{XYZ}=M_{sG}$ gives us the following equation
\begin{equation}
\label{mu1}
4^{2(1-\eta)} \frac{\gamma_r}{4\sin^2(\pi \eta)}= \frac{ -\alpha \gamma_r Ja }{ 2\kappa(\eta)}.
\end{equation}

Using the following property of $\Gamma$ functions, 
\begin{equation}
\Gamma(\eta)\Gamma(1-\eta)= \frac{\pi}{\sin{(\pi\eta)}}
\end{equation}
we obtain

\begin{eqnarray}
\alpha&=&-\frac{4^{2(1-\eta)} \Gamma^2(\eta) }{4\pi^2(1-\eta)}\left(  \frac{\sqrt{\pi}  \Gamma(\frac{1}{2(1-\eta)}) }{2\Gamma(\frac{\eta}{2(1-\eta)}) }  \right)^{2(1-\eta)}\nonumber\\
&=&-\frac{\Gamma^2(\eta) }{4\pi^{1+\eta} 2^{2(\eta-1)}  (1-\eta)}  \left(  \frac{  \Gamma(\frac{1}{2(1-\eta)}) }{\Gamma(\frac{\eta}{2(1-\eta)}) }  \right)^{2(1-\eta)} .
\end{eqnarray}

Note that due to the $U(1)$ symmetry at $\gamma=0$,
\begin{equation}
\sum_r S^{x}_{r}S^{x}_{r+1} = -\sum_r S^{y}_{r}S^{y}_{r+1} \to  \frac{\alpha}{2r_0^{1-2\eta}} \int \mathrm{d} r  \cos{ \sqrt{8 \pi \eta } \Theta}. 
\end{equation}

Also note that at $\gamma=0$ the effective theory enjoys conformal invariance and hence a unique normalization of correlation function is carried to all distances Eq. (\ref{SDN}).   
Finally we obtain for $\gamma=0$,
\begin{equation}
G^{x,x}_u(r) =G^{y,y}_u(r) = \frac{B_2}{r^{4\eta}},
\end{equation}
where 

\begin{eqnarray}
B_2&=& \frac{\alpha^2 }{2^3}= \frac{1}{2^7}\frac{\Gamma^4(\eta) }{\pi^{2+2\eta} 2^{4(\eta-1)}  (1-\eta)^2}  \left(  \frac{  \Gamma(\frac{1}{2(1-\eta)}) }{\Gamma(\frac{\eta}{2(1-\eta)}) }  \right)^{4(1-\eta)}\nonumber\\
&=& \frac{\Gamma^4(\eta) }{\pi^{2+2\eta} 2^{3+4\eta}  (1-\eta)^2}  \left(  \frac{  \Gamma(\frac{1}{2(1-\eta)}) }{\Gamma(\frac{\eta}{2(1-\eta)}) }  \right)^{4(1-\eta)}.
\end{eqnarray}
This expression agrees with the one obtained in [\onlinecite{LT}].

We note that one cannot use the effective representation of single-spin operators \cite{GNT} to obtain the short-distance correlation function for the XXZ spin-$\frac{1}{2}$ chain and in particular to obtain exact amplitudes of the correlation functions of bilinear spin operators with the fusion rules of underlying conformal theory. This is so, because conformal symmetry is only an effective property of the model and at short distances the XXZ chain is not conformally invariant, because of irrelevant, in infrared limit, corrections (the leading ones can be found in [\onlinecite{Lukyanov98}]). Due to this reason the constant $B_2$ is not related to coefficients $A^x_0$ and $A^x_1$ appearing in Eq. (3) of the main text in any simple way and also we cannot determine the exact numerical prefactor in front of the staggered part of the bilinear spin correlation function [the $B_1$ coefficient in Eq. (7) of the main text cannot be fixed with currently known methods].

\subsection{Calculating anomalous dimension $\gamma_{B{_2}}$}

In this appendix we show how to calculate the anomalous dimension of the field $\cos {\sqrt{8 \pi } \Theta}$ picked up upon renormalization due to marginally irrelevant perturbations of the Gaussian model. 

In the absence of perturbations, for $\vec g=0$, the effective theory given by Eq. (4) in the main text has conformal invariance, and hence 
\begin{equation}
\label{cw}
2\langle  \cos {\sqrt{8 \pi } \Theta(0)} \cos {\sqrt{8 \pi } \Theta (r)} \rangle_G =r^{-4}.
\end{equation}
When marginally irrelevant perturbations are included on top of the Gaussian model, $g_{\bot}$ does not contribute to the anomalous dimension of the $\cos {\sqrt{8 \pi } \Theta}$ field to first order, since
\begin{equation}
\langle  \cos {\sqrt{8 \pi } \Theta(0)} \int \mathrm{d^2} {\vec x} \mathcal{H'}_2(\vec x)  \cos {\sqrt{8 \pi } \Theta (r)} \rangle_G =0.
\end{equation}
Hence, at the lowest (first) order in $\vec g\neq 0$, we can include $g_{||}$ into the quadratic part of the action and obtain
\begin{equation}
\label{andimension}
G^{x,x}_{B_2}(r) \sim {r^{-2 ( 2-4\pi g_{||} / \sqrt{3})}}.
\end{equation}
Perturbation $\mathcal{H'}_1$ can be included into the quadratic part of the action independently of the strength of $g_{||}$. It is the strength of $g_{\bot}$ that must be small in order to use the anomalous dimension obtained from perturbative analyses at the lowest order.

From Eq. (\ref{andimension}) we read off the anomalous dimension of the $\cos {\sqrt{8 \pi } \Theta}$ field at the lowest order in $\vec g$,
\begin{equation}
\gamma_{B_2}(\vec g)= 2-4\pi g_{||}/\sqrt{3}. 
\end{equation}
Using the fixed-point value of $g_{||}(\infty)=\sqrt{3}(1- \eta)/(2\pi )$ in Eq. (\ref{andimension}) reproduces the $r$ dependence of $G^{x,x}_{B_2}$ in Eq. (7) of the main text, $\sim r^{-4\eta}$.

Next we provide the details of calculating the exact long-distance asymptotics of $G^{x,x}_{B_2}$ at the $SU(2)$ antiferromagnetic point, given in Eq. (14) of the main text. 

\subsection{RG improved perturbation theory approach for long-distance asymptotics of $G^{x,x}_{B_2}$ }

Here we will generalize the calculation of exact asymptotic correlation functions of single-spin operators at the $SU(2)$ antiferromagnetic point \cite{Affleck98,Lukyanov98} to the case of $G^{x,x}_{B_2}$.

Our aim is to compute the two-point correlation function for the effective action with the bare coupling constants $\vec g^0$ (which carry information of the initial microscopic lattice model) $G(r)=G(r,r_0, \vec g(r_0))$. However, since the Hamiltonian is not Gaussian, one has to use some approximate methods for computing correlation functions. If one tries to perform a perturbation theory calculation in coupling constants, a standard method of interacting field theory, because of the logarithmic divergences that occur in the infrared limit, one cannot stop perturbative series at some finite order, even if initially $\vec g(r_0) \ll 1$. For example, in our case, the first order in the coupling constants contribution in the correlation function $G_{B_2}(r)$ comes with $g_{||}$,
\begin{eqnarray}
-2\langle  \cos {\sqrt{8 \pi } \Theta(0)} \int \mathrm{d^2} {\vec x} \mathcal{H'}_1(\vec x)  \cos {\sqrt{8 \pi } \Theta (r)} \rangle_G \nonumber\\
=\frac{8\pi g^0_{||}}{\sqrt{3}} \ln{(r/r_0)}r^{-4}.
\end{eqnarray}

Combining this correction with Eq. (\ref{cw}) we obtain, up to the first order in coupling constants,
\begin{equation}
\label{cw1}
G^{x,x}_{B_2}(r) \sim r^{-4}( 1+\frac{8\pi g^0_{||}}{\sqrt{3}} \ln{r/r_0}).
\end{equation}
Hence, the effective expansion parameter of perturbation series increases logarithmically at large distances, $g^0_{||}\to  g^0_{||}  \ln{(r/r_0)}$.

RG is a way to resum the leading logarithmic divergences of the infinite perturbation series occurring in the $r\to \infty$ limit. One can obtain from Eq. (\ref{cw1}) directly at the $SU(2)$ antiferromagnetic point the double-logarithmic correction of the correlation function as follows. At $\eta=1$ we have $g_{\bot}=g_{||}=g$ and considering it as a small perturbation the following connection between the bare and renormalized couplings exists from the one-loop beta function,
\begin{equation}
\label{KTSU2}
g(r)=\frac{g^0}{1+4g^0 \ln(r/r_0)/\sqrt{3}}.
\end{equation}
Hence to the lowest order in coupling constant we can make a substitution, 
\begin{equation}
\label{subst}
 1+\frac{8\pi g^0}{\sqrt{3}} \ln{\frac{r}{r_0}}= (g^0/g(r))^2+\cdots
\end{equation}
and represent Eq. (\ref{cw1}) in the following form,
\begin{equation}
\label{cw2}
G^{x,x}_{B_2}(r)  \sim {r}^{-4}(g^0/g(r))^2+\cdots.
\end{equation}
From Eq. (\ref{KTSU2}), at large distances, $g(r) \simeq \sqrt{3}/(4 \ln(r/r_0))$, and plugging this into Eq. (\ref{cw2}) produces multiplicative double-logarithmic correction of the algebraic $1/r^4$ decay of the correlation function $G^{x,x}_{B_2}(r)$. Note that if the anomalous dimension of the operator does not depend on the coupling constants (which is the case for conserved quantities) there will be no multiplicative logarithmic corrections in the corresponding correlation function. 

Moreover, apart from the logarithmic correction we can even determine the precise numerical prefactor, by comparing with the exact results for $\eta<1$ \cite{LT}. Since in the infrared limit the running coupling constant $g_{\perp}$ flows to zero, one can estimate the (renormalized) correlation function at large scale, from perturbative expansion in $g_{\perp}$. 

We note that $G^{x,x}_{B_2}$ obeys the following Callan-Symanzik (CS) RG equation,
\begin{equation}
\label{RGEQ}
\!\Big(\frac{\partial}  {\partial \ln \tilde r} +\!\!\!\!\sum_{j=\{||,\bot \}}\!\!\!\!\! \beta_j\frac{ \partial}{ \partial g_j}-2\gamma_{B_2}(\vec g)\Big) G^{x,x}_{B_2}(r,\tilde r, \vec g(\tilde r)) =0,
\end{equation}
where $\vec g(\tilde r)=(g_{||}(\tilde r), g_{\bot}(\tilde r))$, $\beta_j$ are their beta functions presented in Eq. (5) of the main text and $\gamma_{B_2}(\vec g)=2-4\pi g_{||}/\sqrt{3}$ is the anomalous dimension of $\hat O^{x}_2$ calculated in the previous section. The CS RG equation is equivalent to the one presented in the main text Eq. (13), up to the sign in front of the anomalous dimension $\gamma_{B_2}$ (due to the fact that increase of the short-distance cutoff is equivalent to decreasing the distances measured in units of the new cutoff).

CS Eq. (\ref{RGEQ}) defines the evolution of the two-point correlation function $G$ under variation of the length scale $\tilde r$ at which the theory is defined. Since the effective theory is derived from the original microscopic lattice model, the initial length scale is given by the lattice constant $r_0$ and is increased in the RG process of gradually eliminating high-energy degrees of freedom.

The following connection between the bare and renormalized correlation functions is provided by the CS Eq. (\ref{RGEQ}), 
\begin{equation}
\label{BR}
G(r,r_0,\vec g(r_0))= G(r,r_1,\vec g(r_1))e^{-2\int^{r_1}_{r_0}  \gamma(\vec g(\tilde r))\mathrm{d}\ln{ \tilde r} }.
\end{equation}
To see this, observe that the left-hand side of Eq. (\ref{BR}) does not depend on some arbitrary scale $r_1$. Applying $r_1\frac{\delta}{\delta r_1}$ to both sides of Eq. (\ref{BR}) reproduces Eq. (\ref{RGEQ}) for $r_1=\tilde r$. 

We will choose $r_1$ large enough, so that $G(r,r_1, \vec g(r_1))$ can be expanded in powers of $g_{\perp}(r_1)\ll 1$. This step is called the RG improvement of the perturbation theory. The zeroth-order term, evaluated by Gaussian fixed point action, with rescaled cutoff $r_1$ is
\begin{equation}
\label{GFPA}
G(r, r_1, \vec g(r_1))|_{g_{\perp}=0}=Const.\left(\frac{r_1}{r r_0}\right)^{4\eta}.
\end{equation}
Note that increasing cutoff from $r_0$ to $r_1$ is equivalent to decreasing distance (measured in new units) by the same factor, $r\to rr_0/r_1$.

Since at the lowest order $\gamma_{B_2}(\vec g)=2-4\pi g_{||}/\sqrt{3}$, we need the solution of the Kosterlitz-Thouless RG equations (5), presented in the main text, only for $g_{||}$,
\begin{equation}
\label{inteq}
g_{||}(\tilde r)={\sqrt{3}(1-\eta)} \coth{\left(2(1-\eta)\ln{\tilde r}\right)}/({2\pi}).
\end{equation} 

Using Eq. (\ref{inteq}) and Eq. (\ref{GFPA}) we get from Eq. (\ref{BR})
\begin{eqnarray}
\label{ppl}
G^{x,x}_{B_2}(r)&=&Const.   \left (  \frac{1}{r}  \right )^{4\eta} \left (  \frac{1-(Ar)^{-4(1-\eta)} }{1-r^{-4(1-\eta)}_0}  \right )^{2} .
\end{eqnarray}
In obtaining Eq. (\ref{ppl}) we used the following table integral $\int \mathrm {d}x \coth{\alpha x} =\ln( \sinh(\alpha x))/\alpha$ and put $r_1=Ar$. As an artifact of the finite order perturbation theory approximation (for correlation function, beta functions and anomalous dimension), $G^{x,x}_{B_2}(r)$ in Eq. (\ref{ppl}) contains some arbitrary number $A$.

Choosing the ($\eta$ dependent) normalization constant in such a way that the leading behavior of $G^{x,x}_{B_2}(r)$, for distances $ \ln{r}\gg 1/(1-\eta)$, becomes identical to that shown in Eq. (11) of the main text, we obtain for $\eta\to 1$ over an intermediate range of distances $1\ll \ln{r}\ll1/(1-\eta)$,
\begin{equation}
\label{pl}
G^{x,x}_{B_2}(r)\simeq {B_2 (4(1-\eta)\ln{(Ar)})^2}/{r^4},
\end{equation}
where for $\eta \to 1$
\begin{equation*}
 B_2\simeq \frac{1-2(1-\eta)\ln{\frac{(1-\eta)}{6\pi}}+O((1-\eta)^2) }{ (1-\eta)^2 2^7 \pi^4 }.
\end{equation*}
For obtaining Eq. (\ref{pl}) from Eq. (\ref{ppl}) we have used the following equation, 
\begin{equation}
\lim_{\eta \to 1}(r^{2(1-\eta)}- r^{-2(1-\eta)})=4(1-\eta) \ln{r}+O((1-\eta)^3).
\end{equation}

Number $A$ in Eq. (\ref{pl}) can be estimated by going to higher order approximation in perturbation theory \cite{LT,Lukyanov98}; however it will contain some fitting constant depending on the order of truncation of the perturbative series. If the complete perturbation series could be summed then the correlation function naturally would not contain any fitting parameter.

\subsection{Ground state fidelity susceptibility at the Berezinskii-Kosterlitz-Thouless quantum phase transition}

In quantum many-body systems, at zero temperature, phase transitions can be encountered when changing strength $\lambda$ of the certain term in the Hamiltonian,
$\hat{H}=\hat{H}_{0}+\lambda \hat{V}$. In the finite-size computational studies, a quantity that can be sensitive to the rapid change of the ground state is the overlap of two ground states at slightly different values of the parameter $\lambda$, $F(\lambda,\lambda+\delta\lambda)=\langle \psi_0(\lambda)|\psi_0(\lambda+\delta\lambda)\rangle$ \cite{ZanardiPaunkovic06} and is called a ground-state fidelity. 
The ground-state fidelity susceptibility per site (FS) \cite{You+07,ZanardiGiordaCozzini07,VenutiZanardi07} is defined as 
\begin{equation}
\label{FS1}
\chi_{L}=(1/L)\lim_{\delta\lambda\to0} \big[-2\ln
|F(\lambda,\lambda+\delta\lambda)|\big]/(\delta\lambda)^{2}
\end{equation}
and is expected to diverge in the thermodynamic limit at certain quantum phase transitions. 

Calculating numerically FS has been established as an unbiased indicator of quantum phase transitions
\cite{Schwandt+09,Gu10rev}, especially in one-dimensional
systems where a highly accurate numerical calculation of the ground state wave function is
possible due to the well established methods, such as DMRG.

Another attractive feature of FS is that even though it can be computed solely from the ground-state wave functions, it contains information about the matrix element of operator $\hat V$ between the ground state and excited states. Due to this property FS was used to confirm numerically \cite{Vekua2} the analytic prediction for the leading low frequency dependence of the regular part of the dynamical current conductivity in gapless systems \cite{Giamarchi}, the relevant experimentally measurable quantity that is notoriously difficult to compute with other numerical methods.

For a translationally invariant system with a non-degenerate ground state, perturbed by a local operator
$\hat{V}=\partial_{\lambda}\hat{H}=\sum_{r}\hat{V}(r)$,
 the following connection between the FS and the reduced two-point correlation function
$G(r,\tau)=\langle\!\langle  \hat{V}(r,\tau)\hat{V}(0,0)\rangle\!\rangle$ exists \cite{VenutiZanardi07},
\begin{equation}
\label{FSF}
\chi_{L}=\int_{r_0}^{L}dr \int_{0}^{\infty} d\tau\, \tau  G(r,\tau),
\end{equation} 
where the imaginary time dependence is defined by $\hat{V}(r,\tau)=e^{\tau
  \hat{H}} \hat{V}(r) e^{-\tau \hat{H}}$, averages are taken in the ground state
$|\psi_{0}(\lambda)\rangle$, and $r_0$ is the short-distance cutoff.
The FS diverges for $L\to\infty$ as $\chi\propto L^{d+2z-2\Delta_{V}}$, where $\Delta_{V}$ is the dimension of
operator $\hat{V}(x)$ at the critical point and $z$ is the dynamic exponent, if $\Delta_{V}\le z+1/2$. 
These simple scaling arguments show that for the Berezinskii-Kosterlitz-Thouless (BKT) phase transition, $z=1$ and
$\Delta_{V}=2$ (the perturbing operator $\hat V$ is marginal) and FS should not diverge.

The XXZ spin-$\frac{1}{2}$ chain is an example model where the properties of the BKT transition in one-dimensional quantum systems can be extracted in an exact way due to integrability.
It is well known that ground-state energy changes smoothly across the BKT transition (ground-state energy is infinitely differentiable with anisotropy parameter across the BKT transition). 
An interesting question is how the ground-state wave function evolves across the transition: 
Does the overlap of the two ground states that correspond to the parameters arbitrarily close, but located at different sides of the BKT transition ``feel'' the transition.
The answer to this question is encoded in the behavior of FS and the correct behavior of FS across the BKT transition point was uncovered in \cite{Vekua1}: The FS develops a cusp singularity, however stays finite in the thermodynamic limit. This is in contrast to the previously seemingly established results on the divergence of the FS at the BKT transition point \cite{MFYang07,Fjarestad08,Sirker10} that were based on asymptotically exact calculations within the effective Gaussian description of the XXZ chain in the gapless regime $\lambda<1$ and extending the results towards $\lambda=1$. Here we clarify why divergence in FS appears within the Gaussian approximation and show that when correcting the Gaussian model by marginally irrelevant perturbations (a necessary ingredient of the effective theory describing the BKT transition), the FS in the thermodynamic limit becomes finite.

In the following we will represent Hamiltonian (1) of the main text as, $\hat{H}_{XXZ}=\hat{H}_{0}+\lambda \hat{V}$, with $\hat V= \sum_{r} S^{z}_{r}S^{z}_{r+1}$.

Since the effective Gaussian model, given by Eq. (2) in the main text, is quadratic, one can explicitly calculate the fidelity $F(K,K+\delta K)$, where $K=1/(4\pi R^{2})$ is the Luttinger liquid parameter, and obtain for the FS in the thermodynamic limit, $\chi|_{L=\infty} =  (\partial_{\lambda}K)^{2}/(8r_0K^{2})$. Stretching the mapping to the Gaussian model towards the $SU(2)$ limit (which is not justified there), the singular dependence $K(\lambda)$, at $\lambda=1$, leads to the divergence of the FS $\chi\propto (1-\lambda)^{-1}$ \cite{MFYang07,Fjarestad08,Sirker10}. FS is related to the overlaps of the ground states at slightly shifted anisotropy parameters and the ground state contains information on all distances, whereas an effective approach only connects low-energy properties of the microscopic and effective models. However, since the singular contribution in FS of the lattice systems, if such exist, is expected to come from large distances (short distances being regularized by lattice), it seems reasonable to assume that for the purpose of identifying singularity in FS, effective description will be reliable. The problem is that the mapping itself of the XXZ spin-$\frac{1}{2}$  model to the Gaussian theory becomes singular for $\lambda\to 1$. This singularity is encoded as well in the dependence of the Luttinger liquid parameter on the anisotropy of the microscopic model when $\lambda \to 1$. Despite the fact that perturbations to the Gaussian model are marginally irrelevant and they die out in the infrared limit (fixed-point value of $g$ is zero), and thus the fixed-point action is Gaussian, they are crucial to be kept for obtaining correct correlation functions and hence for calculating the FS due to Eq. (\ref{FSF}).

If one does naive extrapolation of Eq. (11) in the main text towards the Heisenberg AFM point, without keeping marginally irrelevant correction to the Gaussian model with the subsequent RG improved procedure outlined above, one obtains unphysical divergence of the prefactor in front of the ${1}/{r^4}$ algebraic decay, $B_2\sim {1}/{(1-\lambda)}$. This would produce an identical erroneous result for the FS that has been obtained by calculating overlaps of the Gaussian model at different Luttinger liquid parameters and extending the result all the way towards $\lambda \to 1$ from the gapless side \cite{MFYang07,Fjarestad08,Sirker10}.

Using non-Abelian bosonization, however, we showed recently that instead of diverging, FS shows a finite (cusplike) peak at the BKT phase transition. It converges, though logarithmically, to its finite thermodynamic value with increasing the system size \cite{Vekua1}, $\chi_L \simeq \chi_{c} - \chi_{1}/\ln(L/a) +\cdots$, where both $\chi_c$ and $\chi_1$ are finite positive numbers that are obtained respectively from the two- and three-point correlation functions of the currents of the $SU_1(2)$ Wess-Zumino model, the fixed-point action of the Heisenberg spin-$\frac{1}{2}$ antiferromagnetic (AFM) chain \cite{GNT}.

To obtain the ground-state FS at the BKT phase transition from Eq. (\ref{FSF}), we need imaginary-time dependence of the uniform part of the bilinear spin correlation function at the $SU(2)$ AFM point. To this end the
effective Lorentz invariance of the Heisenberg spin-$\frac{1}{2}$ chain can be invoked to calculate $G^{z,z}_{u}(r,\tau)$, noting that the processes breaking Lorentz 
invariance (due to lattice) have high scaling dimension \cite{Lukyanov98} and will not modify asymptotic results. We can represent in bosonization $\hat V$ as
\begin{equation}
\!\sum_rS_rS_{r+1} \to \!\alpha_0 H_{eff}+\!\!\int\!(\alpha_1  \mathcal{H'}_1(r)+ \alpha_2\mathcal{H'}_2(r))\mathrm{d}r,
\end{equation}
where $H_{eff},  \mathcal{H'}_1,  \mathcal{H'}_2$ are from Eq. (4) of the main text and proportionality factors $\alpha_0, \alpha_1$, and $\alpha_2$ will not be important in the following. The first term in the right-hand side of bosonization correspondence is the Hamiltonian of the effective model $\hat{H}_{XXZ} \to H_{eff}$; hence it does not contribute to the FS of $\hat V$, like any other quantity that commutes with the Hamiltonian. This is easily seen from the following representation of the ground state FS \cite{You+07},
$
\chi=\sum_{n\neq 0} {|\langle n| \hat V |0\rangle  |}/{(E_n-E_0)^2}.
$
Both $ \mathcal{H'}_1$ and $\mathcal{H'}_2$ are scalar operators and hence for calculating the FS we can obtain the imaginary-time dependence of the correlation function $G^{z,z}_{u}(r,\tau)$ from the equal-time result, Eq. (18) of the main text, by replacing $r\to \sqrt {r^2+v^2\tau^2}$.

From the convergence of the integral at large distances it is clear that FS does not diverge at the BKT phase transition in the thermodynamic limit, $L=\infty$,
\begin{equation}
\chi\sim \int_{r_0}^{\infty}dr \int_{0}^{\infty} d\tau\, \tau \frac{\ln^2{(r^2+v^2\tau^2)}}{(r^2+v^2\tau^2)^2} <\infty.
\end{equation}



\begin{thebibliography}{10}
\bibitem{comment0} In the effective theory correspondence of the bilinear spin operators at $\lambda=1$ no index $a$ appears in the leading term (which is a staggered contribution and is called a dimerization operator \cite{GNT}): $S^{a}_rS^{a}_{r+1}-{e_0}/{3}\to (-1)^r\cos{\sqrt{2\pi}\phi(r)}+\cdots$, where dots indicate subleading contribution, including the uniform components, that depend on index $a$ and hence will contribute to the difference of correlation functions $G^{a,a}_r-G^{a,b\neq a}_r$.
\bibitem{comment1} The uniform part $G^{a,a}_u(r)$ contributes to the weak oscillations between even and odd $r$ data (even-odd effect). With increasing $r$ oscillations die out, in accordance with the effective theory picture. We note that, since $G^{a,a}_u(r)>0$, correlations at distances that are even multiple of the lattice constant get enhanced, as opposed to the case of the sigle-spin correlation functions Eq. (\ref{su2}).
\end{thebibliography}
\end{document}